\documentclass[aps,pre,twocolumn,groupedaddress,superscriptaddress]{revtex4}
\usepackage{graphicx,multirow}
\usepackage{amsmath,dcolumn,txfonts}

\begin{document}

\title{Impact of misinformation in temporal network epidemiology}

\author{Petter Holme}
%\email[E-mail:]{holme@cns.pi.titech.ac.jp}
\affiliation{Institute of Innovative Research, Tokyo Institute of Technology, Tokyo, Japan}
\author{Luis E. C. Rocha}
\affiliation{Department of Public Health Sciences, Karolinska Institutet, Stockholm, Sweden}\affiliation{Department of Mathematics, Universit\'e de Namur, Namur, Belgium}

\begin{abstract}
We investigate the impact of misinformation about the contact structure on the ability to predict disease outbreaks. We base our study on $31$ empirical temporal networks and tune the frequencies in errors in the node identities or timestamps of contacts. We find that for both these spreading scenarios, the maximal misprediction of both the outbreak size and time to extinction follows an stretched exponential convergence as a function of the error frequency. We furthermore determine the temporal-network structural factors influencing the parameters of this convergence.
\end{abstract}

\maketitle

\section{Introduction}

Infectious diseases are a major burden to global health. They spread over temporal networks of human contacts~\cite{masuda_holme_rev,masuda_lambiotte,bansal_etal,read}. The structures of such networks affect the dynamics of disease, so to be able to mitigate outbreaks, we need to understand how this happens~\cite{gies,andersonmay,hethcote,keeling_rev}. With new data sources people have been able to map human contact patters to a much greater precision than ever before~\cite{salathe,stopczynski2014measuring,office,school,hschool,conference,hospital}. There is a growing body of literature using such proximity networks as the underlying structure for simulations of disease spreading~\cite{masuda_holme_rev,salathe,masuda_lambiotte,holme_modern,holme_saramaki}. Typical research questions concern: How to exploit the temporal information in vaccination campaigns~\cite{holme_tempo_vacci,office}. How to identify hot-spots for disease spreading in animal trade~\cite{colizza}. How to reduce temporal networks to static networks as accurately as possible for static-network modeling of disease spreading~\cite{holme_pcb}. As with all empirical data, the proximity network these studies are based on come with inaccuracies. These affect the predictions from simulation studies using them as a substrate. In this work, we investigate the impact of such inaccuracies. We compare outbreak predictions in the presence of misinformation, or noise, both in the topological information (who is in contact with whom) and the temporal information (when these contacts happen).

This work is based on simulations of the susceptible-infected-recovered (SIR) model. This is the canonical model for emerging disease outbreaks of pathogens making the infected immune upon recovery. We will assume the contagion can happen during contacts of temporal networks. We use 31 empirical networks as our input. These have different degree of relevance for modeling disease spreading. Some of them records people being in close proximity and thus in danger of spreading e.g.\ influenza. We also use some other data sets from social media, that of course are less realistic as substrates for disease propagation, but are related to the spread of information. An alternative approach would be to use generative models where the temporal network structure can be controlled. This would have the advantage that one can systematically control one structure, but the drawback that one would have to focus on certain structures (like the time between events~\cite{vazquez}). At the time of writing, it is not completely known what temporal network structures that are the most important for disease spreading~\cite{holme_tempdis}. So instead of building a model upon guesses about that, we use empirical networks, which makes it possible to study both how the sensitivity to noise depends on temporal network structure and the empirical networks \emph{per se}. This approach also enables us to discover common features and fluctuations in the contact data sets.

In our simulations, we scan the entire parameter space of the SIR model. (For temporal networks, the parameter space is two-dimensional, unlike for static network epidemiology where the qualitative behavior is determined by only one parameter.) We compare these disease simulations for the empirical networks with the same type of simulation on networks where the identities of the nodes and the time of the contacts have been altered randomly to mimic errors in the data. Then we proceed to measure the largest deviation between the predictions about the time to extinction and the number of affected nodes in the outbreaks. Finally, we try to relate the magnitude of these deviations to the temporal network structure of the data.

\begin{table*}
\caption{\label{tab:data}Basic statistics of the empirical temporal networks. $N$ is the number of individuals; $C$ is the number of contacts; $T$ is the total sampling time; $\Delta t$ is the time resolution of the data set and $M$ is the number of links in the projected static networks.}
\begin{ruledtabular}
\begin{tabular}{l|dddddl}
Data set & N & C & T & \Delta t & M & Ref.\\ \hline
\textit{Conference} & 113 & 20,818 & 2.50d & 20s & 2,196 & \cite{conference}\\
\textit{Hospital} & 75 & 32,424 & 96.5h & 20s & 1,139 & \cite{hospital} \\
\textit{Office} & 92 & 9,827 & 11.4d & 20s & 755 & \cite{office} \\
\textit{Primary School 1} & 236 & 60,623 & 8.64h & 20s & 5,901 & \cite{school} \\
\textit{Primary School 2} & 238 & 65,150 & 8.58h & 20s & 5,541 & \cite{school} \\
\textit{High School 1} & 312 & 28,780 & 4.99h & 20s & 2,242 & \cite{hschool} \\
\textit{High School 2} & 310 & 47,338 & 8.99h & 20s & 2,573 & \cite{hschool} \\
\textit{High School 3} & 303 & 40,174 & 8.99h & 20s & 2,161 & \cite{hschool} \\
\textit{High School 4} & 295 & 37,279 & 8.99h & 20s & 2,162 & \cite{hschool} \\
\textit{High School 5} & 299 & 34,937 & 8.99h & 20s & 2,075 & \cite{hschool} \\
\textit{Gallery 1} & 200 & 5,943 & 7.80h & 20s & 714 & \cite{gallery} \\
\textit{Gallery 2} & 204 & 6,709 & 8.05h & 20s & 739 & \cite{gallery} \\
\textit{Gallery 3} & 186 & 5,691 & 7.39h & 20s & 615 & \cite{gallery} \\
\textit{Gallery 4} & 211 & 7,409 & 8.01h & 20s & 563 & \cite{gallery} \\
\textit{Gallery 5} & 215 & 7,634 & 5.61h & 20s & 967 & \cite{gallery} \\
\textit{Reality} & 64 & 26,260 & 8.63h & 5s & 722 & \cite{reality} \\
\textit{Romania} & 42 & 1,748,401 & 62.8d & 1m & 256 & \cite{roman} \\
\textit{Kenya} & 52 & 2,070 & 61h & 1h & 86 & \cite{kenya} \\
\textit{Diary} & 49 & 2,143 & 418d & 1d & 345 & \cite{read} \\
\textit{Prostitution} & 16,730 & 50,632 & 6.00y & 1d & 39,044 & \cite{prostitution} \\
\textit{WiFi} & 18,719 & 9,094,619 & 83.7d & 5m & 884,800 & \cite{wifi} \\
\textit{UK} & 25 & 408,996 & 74d & 1s & 139 & \cite{dsn} \\
\textit{Messages} & 35,624 & 489,653 & 3,018d & 1s & 94,768 & \cite{karimi}  \\
\textit{Forum} & 7,084 & 1,429,573 & 3,141d & 1s & 138,144 & \cite{karimi} \\
\textit{Dating} & 29,341 & 529,890 & 512d & 1s & 115,684 & \cite{pok} \\
\textit{College} & 1,899 & 59,835 & 193d & 1s & 13,838 & \cite{college} \\
\textit{Facebook} & 45,813 & 855,542 & 1,561d & 1s & 183,412 & \cite{mislove} \\
\textit{E-mail 1} & 57,194 & 444,160 & 112d & 1s & 92,442 & \cite{ebel} \\
\textit{E-mail 2} & 3,188 & 309,125 & 81d & 1s & 31,857 & \cite{eckmann} \\
\textit{E-mail 3} & 986 & 332,334 & 526d & 1s & 16,064 & \cite{eml3} \\
\textit{E-mail 4} & 167 & 82,927 & 271d & 1s & 3,251 & \cite{radek} \\
\end{tabular}
\end{ruledtabular}
\end{table*}

\section{Preliminaries}

In this section, we will go through some technicalities of our simulation study.

\subsection{Definitions}

 We represent a temporal network $G$ as \textit{contact sequence}---a list of triples $(i,j,t)$ recording a \textit{contact} between $i$ and $j$ at time $t$~\cite{holme_saramaki,holme_modern}. We call a pair of nodes with at least one contact a \textit{link}. We use $N$ and $C$ to represent the number of nodes and contacts, while $T$ represents the \textit{duration} of the temporal network (the time between the first and last contact). Without loss of generality, we can identify the nodes with numbers in the interval $[1,N]$.

\subsection{Contact networks}

As motivated in the Introduction, we base our study on empirical temporal networks. The first class of such networks---and the one most relevant to disease spreading---is human proximity networks. These are data sets that capture when two persons are in close proximity. Many of these data sets come from the Sociopatterns project (sociopatterns.org). These networks are based on people wearing radio-frequency identification sensors that detect contacts between people within 1--1.5 m. One of these datasets comes from a conference~\cite{conference} (\textit{Conference}), another from a school (\textit{Primary School})~\cite{school}, a third from a hospital (\textit{Hospital})~\cite{hospital}, a fourth from an art gallery (\textit{Gallery})~\cite{gallery}, a fifth from office (\textit{Office})~\cite{office}, and a sixth from members of five families in rural \textit{Kenya}~\cite{kenya}. 
The \textit{Gallery} data sets consists of several days where we use the first five. Yet a data set \textit{Reality} was gathered using the Bluetooth channel of the phones of college students~\cite{reality}. In \textit{Romania}, the WiFi channel of smartphones was used to log the proximity between university students~\cite{roman}. \textit{UK} is another similar dataset of the proximity of university students from wearable sensors~\cite{dsn}. The final proximity data, the \textit{Prostitution} network, comes from from self-reported sexual contacts between female sex-workers and their male clients~\cite{prostitution}. This is a special form of proximity network since the contacts represent more than just proximity (i.e.\ sexual contacts).

In addition to the proximity networks, we also study networks from electronic communication. \textit{Facebook} comes from the wall posts at the social media platform Facebook~\cite{mislove}. \textit{College} records the network of communication at a Facebook-like service~\cite{college}. \textit{Dating} gives the interaction at an early Internet dating website~\cite{pok}. \textit{Messages} and \textit{Forum} are records of user interaction at a film community~\cite{karimi}. Finally we use two data sets of e-mail communication. One, \textit{E-mail 1}, recording all e-mails to and from a set of sampled accounts~\cite{ebel}. The other three, \textit{E-mail 2}~\cite{eckmann}, \textit{3}~\cite{eml3}, and \textit{4}~\cite{radek} recording e-mails within a set of sampled accounts. We list basic statistics---sizes, sampling durations, etc.---of all the data sets in Table~\ref{tab:data}.

\subsection{Epidemic simulation}

There are a few different ways to simulate SIR dynamics on temporal networks. We use the following approach. First, we set all individuals to S (susceptible). Then, we randomly choose one node $i_0\in[1,N]$ and one time $t_0\in[0,T)$ and change $i_0$ from S to I at time $t_0$. Then we go through the contacts temporal network by order of their time stamp. If a contact connects a susceptible and an infected node, the susceptible can become infected with probability $\lambda$. An infectious stays infectious $\delta$ time steps before becoming recovered. When there are no infectious individuals and time is later than $t_0$ the outbreak is considered extinct. Note that this definition is slightly different from the more common one~\cite{hethcote} (where an infectious individual have the same chance of getting well every time step), but could be motivated by being algorithmically simpler~\cite{holme_versions} and not less realistic~\cite{olga}. We characterize an outbreak by the time to extinction $\tau T$ and the average outbreak size $\Omega N$. $\tau$ and $\Omega$ are thus  quantities normalized to the interval $[0,1]$. For every temporal network, we average over $10^5$ runs of the disease simulation.

\subsection{Controlling misinformation}

To model errors in the temporal information, we replace the time stamps of a fraction $\epsilon_T$ of the contacts of $G$ by random times in the interval $[1,T]$. Similarly, for investigating the response to the information about node identities, we replace a randomly selected fraction $\epsilon_N$ of the node id-numbers by random numbers in the interval $[1,N]$. The only two constraints we impose in these randomization schemes is that the resulting contacts should not introduce multiple links or self-links. If a generated node-id number does not satisfy the constraint, we redraw it. We do not study misinformation in both time and node identities---if $\epsilon_T>0$ then $\epsilon_N=0$ and vice versa.

Technically, this approach is similar to randomization techniques~\cite{holme2005,Karsai2011} where the temporal network structure is investigated by systematically randomizing away some structure---like the order of contacts---and studying the response to quantities characterizing the functionality of the network (like $\Omega$, $\tau$, etc.). The difference is that we tune the randomization via $\epsilon$ to monitor the response.

\subsection{Measuring sensitivity to misinformation}

We use two quantities to characterize the severity of an outbreak: the average final outbreak size $\Omega$---the fraction of the population that are R after the outbreak---and the extinction time $\tau$---the time from the first to the last infected individual in the population. Let
\begin{equation}\label{eq:delta}
\Delta(\epsilon,\lambda,\delta)=\langle\Omega(G_{\epsilon},\lambda,\delta)\rangle - \Omega(G,\lambda,\delta),
\end{equation}
where $\langle\:\cdot\:\rangle$ denotes the average over an ensemble of networks $G_\epsilon$ in which a fraction $\epsilon$ of misinformation has been imposed to the node identities and $G$ is the original network. Analogously, we define $\Delta\Omega$ for the deviation of outbreak sizes.

To study the response of $\Delta$ to noise across the SIR parameter space, we use the maximal deviation
\begin{equation}\label{eq:epsilon}
\omega(\epsilon)=\max\Big[ \max_{\delta,\lambda}\Delta(\epsilon,\delta,\lambda),-\min_{\delta,\lambda}\Delta(\epsilon,\delta,\lambda)\Big],
\end{equation}
where $\Delta$ can represent both $\Delta\Omega$ and $\Delta\tau$.  Furthermore, we drop the $N$ or $T$ subscripts of $\omega$ and $\epsilon$. With a specific disease and a specific network in mind, one should of course investigate its feasible region of the $\lambda,\delta$ space.

\subsection{Network descriptors}
\label{net_desc}
How much misinformation impacts prediction of $\Omega$ and $\tau$ depends in the the structure of the temporal networks. To understand this, we consider 48 quantities, or \textit{network descriptors}. Refs.~\cite{holme_masuda_r0,holme_tempdis} use a similar approach but a different set of measures.

\subsubsection{Long-term activity}

A first type of network descriptors characterizes the long-term activity of the data---about if the overall activity is constant throughout the sampling time, if the links are active throughout the sampling period, etc. $x_{nT}$ and $x_{lT}$ show how large fraction of the nodes and links, respectively, that are present at half the duration of the data. In a situation where individuals are participating in contacts already early in the data, $x_{nT}$ will be small. It will be large if nodes enter the data throughout the sampling period. Analogously, $x_{lT}$ will be small if the contacts between all nodes that have at least one contact happen early in the sampling period. Some of the data sets have a more intense activity at the end of the sampling period. To compensate for this, we also measure $x_{nC}$ and $x_{lC}$---the corresponding quantities to $x_{nT}$ and $x_{lT}$ but measured at the time half of the contacts have been observed. A fifth final measure of this type is $x_C$---the fraction of contacts observed at half the sampling time. In a data set with a growing level of activity, this quantity would be less than $1/2$.

\subsubsection{Durations of nodes and links}

Our second type of network descriptors relate to the life-span of nodes and links in the data. We start from the set of durations of nodes (links) presence in the data (normalized by $T$), i.e.\ the time between the first and last contact a node (link) participates in. Then we use four summary statistics to describe these sets---the mean $\mu^d_n$ ($\mu^d_l$), standard deviation $\sigma^d_n$ ($\sigma^d_l$), coefficient of variation $v^d_n=\sigma^d_n/\mu^d_n$ ($v^d_l=\sigma^d_l/\mu^d_l$) and finally the skewness $\gamma^d_n$ ($\gamma^d_l$). $\gamma=\mu_3/\mu_2^{3/2}$ where $\mu_2$ and $\mu_3$ are the second and third moments of the distribution, respectively. A long enough temporal network where most of the nodes are present throughout the duration of the data, the average $\mu^d_n$ would be relatively large.

\subsubsection{Inter-event times of nodes and links}

The times between contacts for individual nodes and links are called \textit{inter-event times}. These are known to be broadly distributed~\cite{goh} which (if it is the only temporal structure present) tend to slow down spreading phenomena~\cite{vazquez}, at least at an early stage of an outbreak~\cite{rocha_blondel}. We use the same summary statistics as for the duration of nodes and links---i.e.\ the mean $\mu^i_n$ ($\mu^i_l$), standard deviation $\sigma^i_n$ ($\sigma^i_l$), coefficient of variation $v^i_n$ ($v^i_l$)---called \textit{burstiness} in Ref.~\cite{goh}---and skewness $\gamma^i_n$ ($\gamma^i_l$).

\subsubsection{Nodes and links activity}

We also study the overall activity in the data---the total number of contacts---of nodes and links. From the sequence of numbers of contacts we calculate the same four summary statistics as for inter-event times and durations. Note that the node and links activities is sometimes referred to as ``strength''~\cite{Onnela01052007}.

\subsubsection{Static network structures}

So far, the quantities mentioned have all concerned temporal aspects of the data in one way or another. We also measure some properties of static networks derived from the contacts. There are several ways to reduce the contacts of a temporal network to a static network~\cite{holme_pcb}. We use two simple methods. First, we consider the network of links between pairs of nodes that have at least one contact in the data. Second, we construct a \textit{reduced network} of pairs of nodes with at least $n$ contacts, where $n$ is chosen as large as possible with the constraint that the size of the largest connected component should be at least a fraction $\alpha$ of its original size. We use $\alpha=0.8$---one do not want it so small that the network is completely fragmented, and not so large that the network is just the same as the network of all links (that we anyway also consider).

For both the two above types of static networks, we measure two classes of network quantities. First, those related to the degree distribution (the same four characterizing the inter-event time and duration distributions---mean, standard deviation, coefficient of variation and skewness). Second, we measure three classic network quantities---the degree assortativity, the clustering coefficient and the number of connected components. The \textit{degree assortativity} is roughly speaking the Pearson correlation of the degree of the nodes connected by a link. Thus it measures the tendency for high-degree nodes to form links with other high-degree nodes, and low-degree to low-degree nodes. The \textit{clustering coefficient} measures the fraction of triangles to the numbers of triples of nodes connected by either two or three links. For an introduction to these types of measures, see text books like Refs.~\cite{barabasi:book,newman:book}.
% * <pttrhlm@gmail.com> 2017-06-17T05:34:32.931Z:
%
% ^.

\subsection{System sizes and summary}

As a final category of temporal network structural measures we will also use the systems sizes---the number of nodes, contacts and links (pairs of nodes having at least one contact). All together, we have twelve categories of network structural measures: long-term activity (with five measures), inter-event times, activities and durations of nodes and links (each with four measures), degree distributions and other network quantities for the full static network and the reduced network (each with four measures), and the above-mentioned system sizes (three measures).

\begin{figure}
\includegraphics[width=0.85\columnwidth]{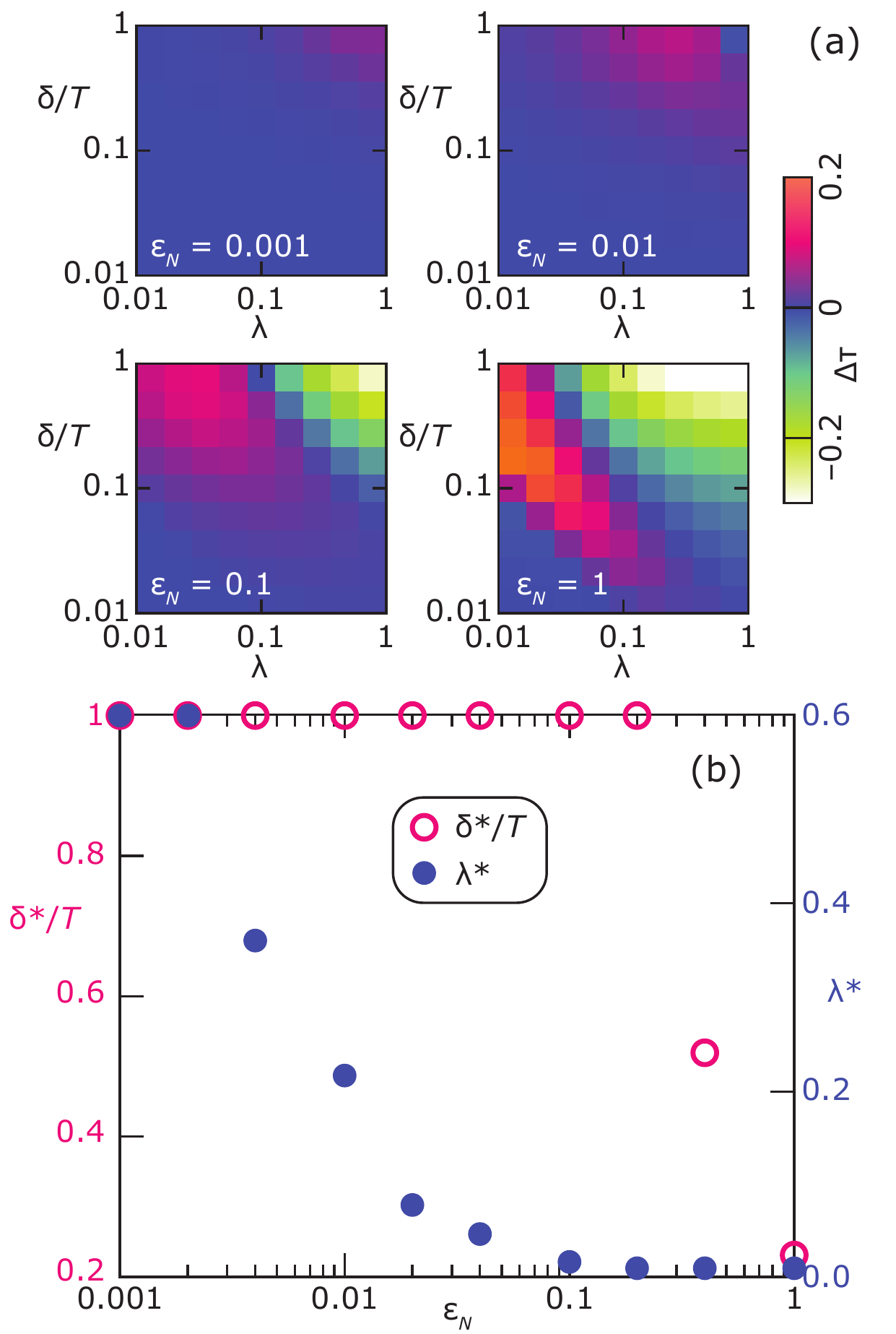}
\caption{Panel (a) shows the deviation of the predicted extinction time $\Delta\tau$ for the \textit{E-mail 4} data set and node-identity error frequencies $\epsilon_N = 0.001$, $0.01$, $0.1$ and $1$ respectively. Panel (b) shows how the location of the maximal $\Delta\tau$ in the SIR parameter space depends on the node error rate.}
\label{fig:eml4_n_xtme}
\end{figure}

\begin{figure*}
\includegraphics[width=0.7\textwidth]{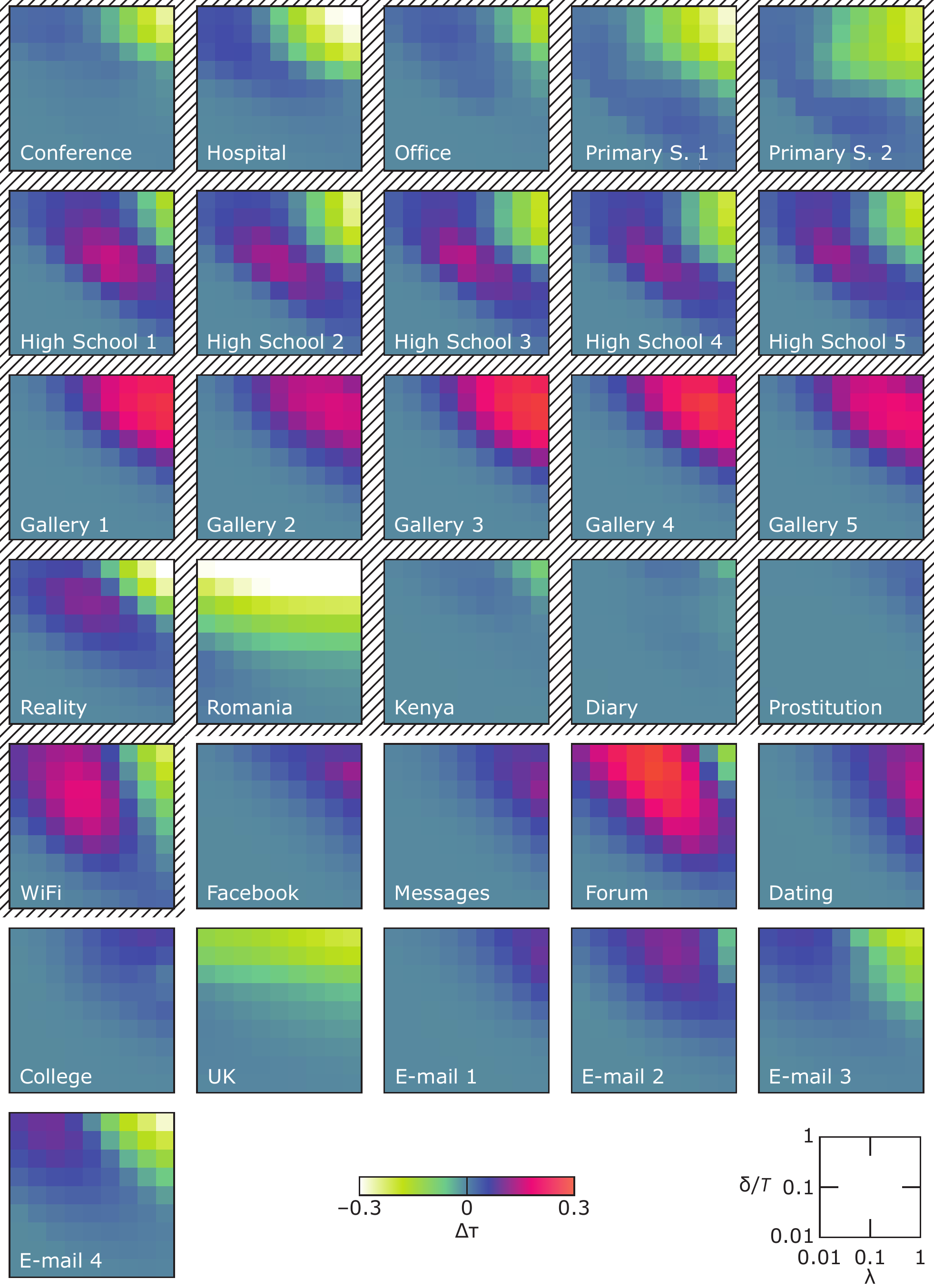}
\caption{ $\Delta\tau$ (the deviation of the predicted extinction time) for all data sets as a function of the SIR parameter values (per-contact transmission probability $\lambda$ and disease duration $\delta$ normalized by the total sampling time $T$). These plots show data for node-identity errors at a level $\epsilon_N=0.1$. The shaded area indicate the data sets of human proximity, the others come from social media and digital communication.}
\label{fig:n_xtme}
\end{figure*}

\begin{figure*}
\includegraphics[width=0.7\textwidth]{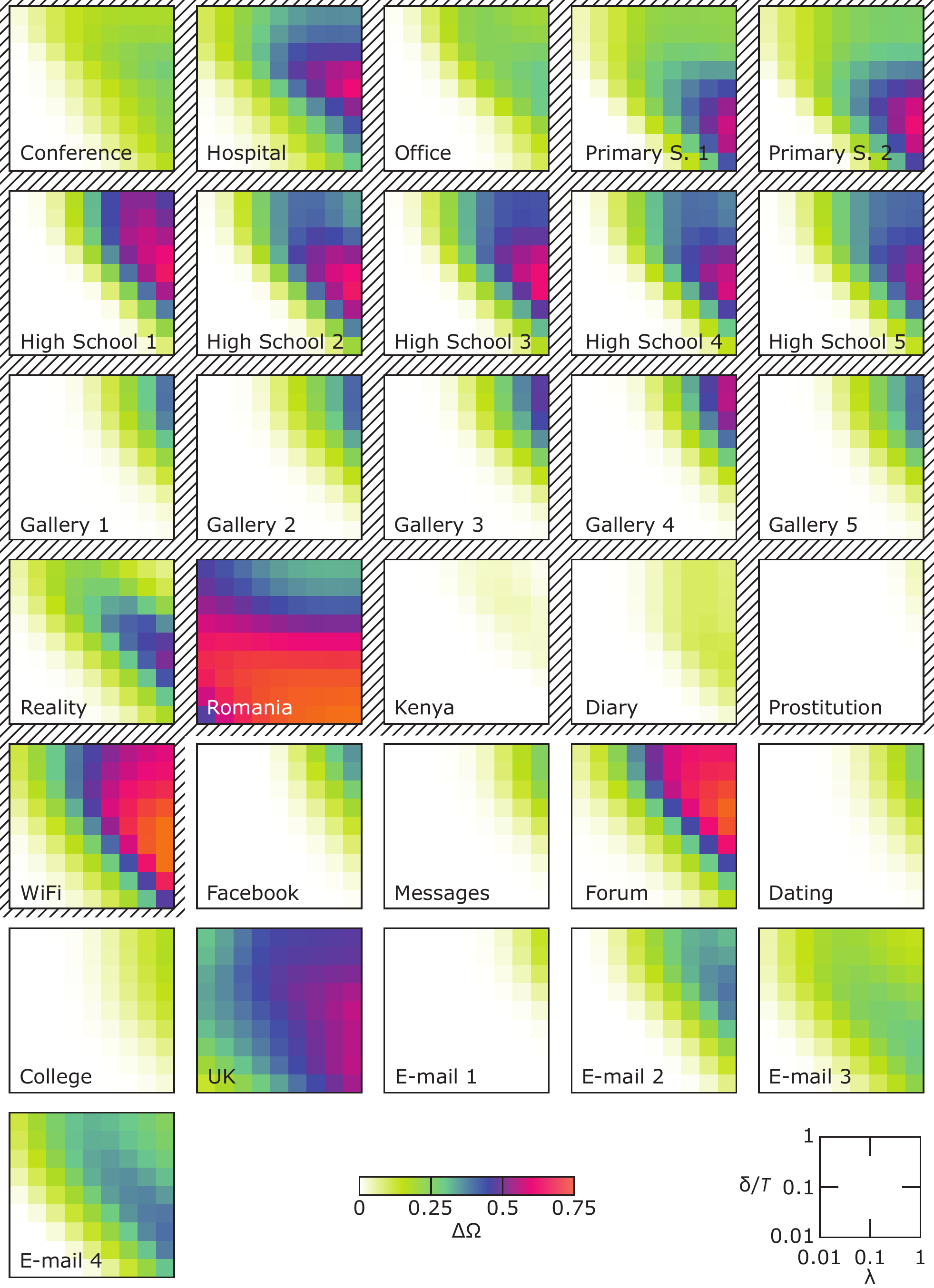}
\caption{$\Delta\Omega$ (the deviation of the predicted outbreak size) for all data sets as a function of the SIR parameter values. The node-identity misinformation is $\epsilon_N=0.1$. The shaded area indicate the human proximity data.}
\label{fig:n_osze}
\end{figure*}

\begin{figure*}
\includegraphics[width=0.7\textwidth]{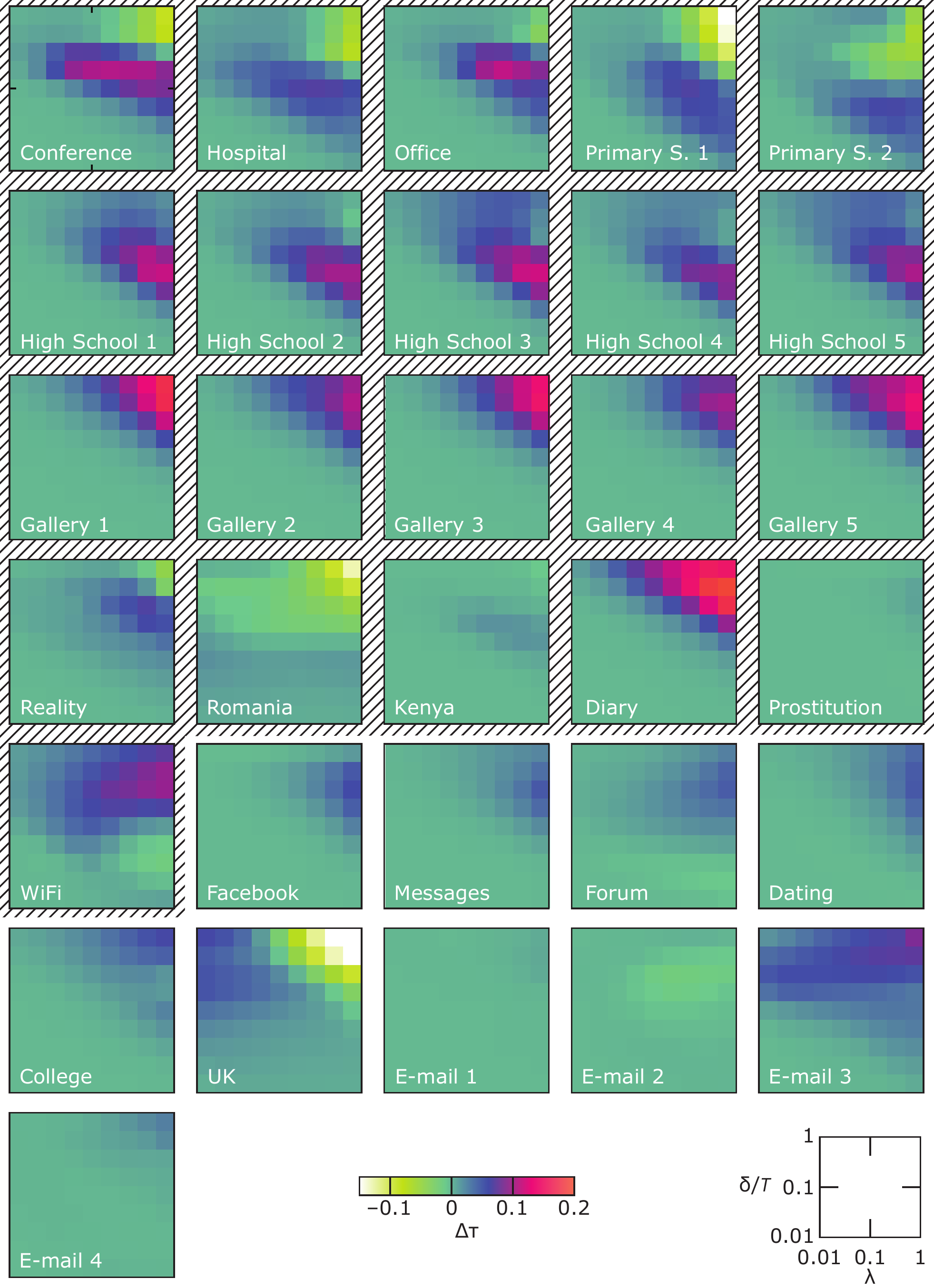}
\caption{ Plots corresponding to Fig.~\ref{fig:n_xtme} but for time-stamp errors at a level $\epsilon_T=0.1$.}
\label{fig:t_xtme}
\end{figure*}

\begin{figure*}
\includegraphics[width=0.7\textwidth]{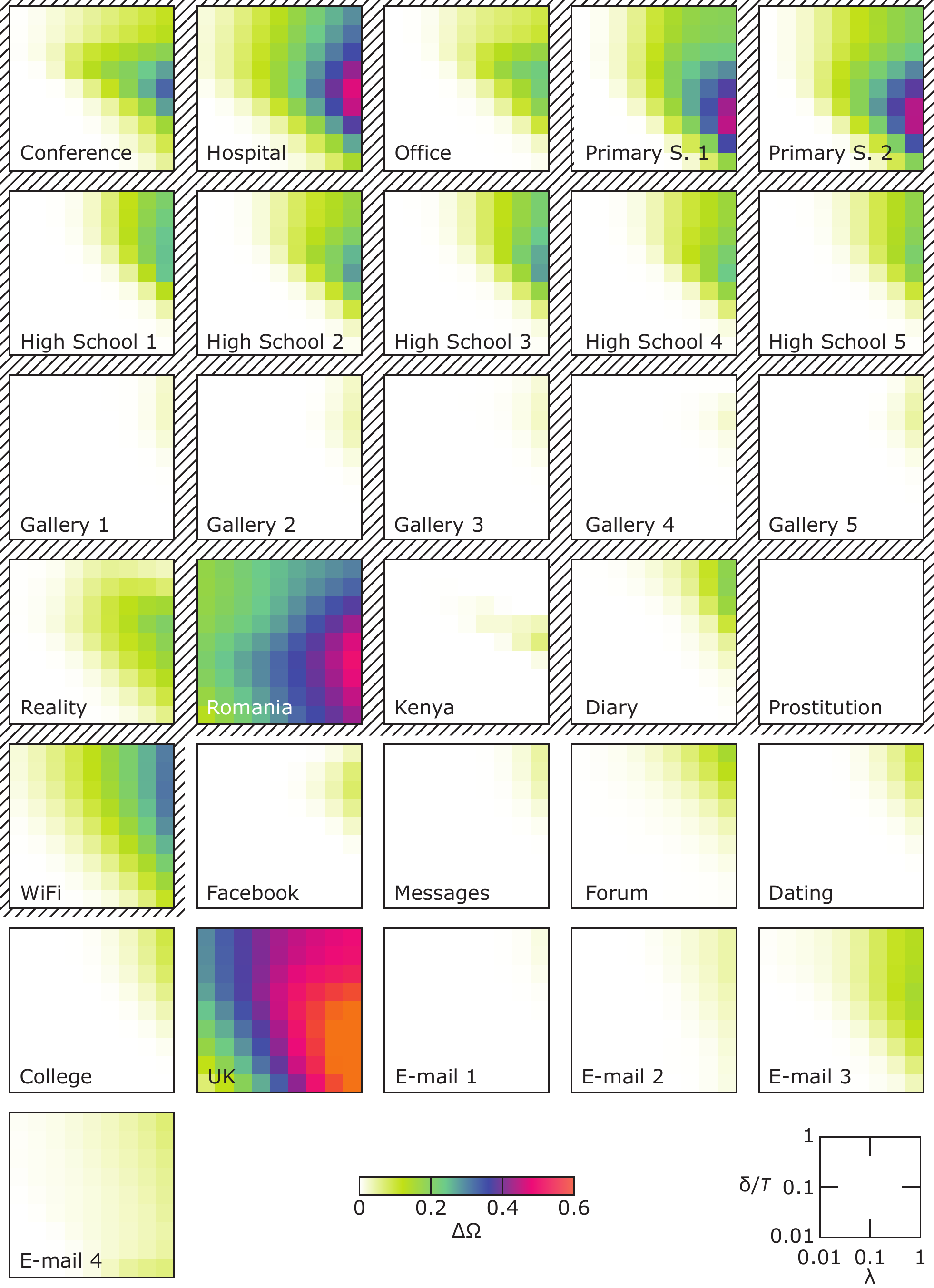}
\caption{ Plots corresponding to Fig.~\ref{fig:n_osze} but for time-stamp errors at a level $\epsilon_T=0.1$.}
\label{fig:t_osze}
\end{figure*}

\begin{figure}
\includegraphics[width=0.55\columnwidth]{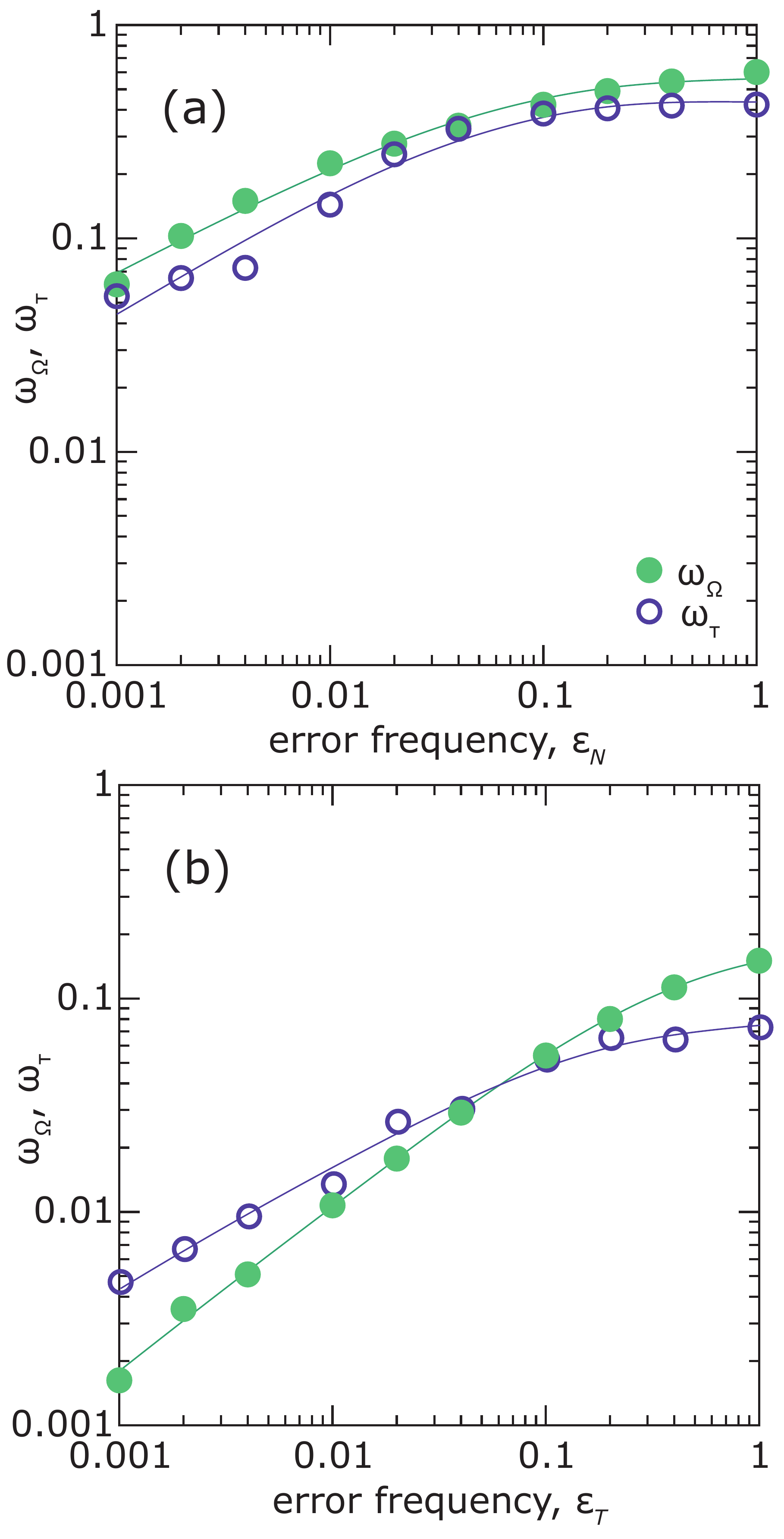}
\caption{ $\omega_\Omega$---the difference between the largest and smallest $\Delta\Omega$ values---for the \textit{E-mail 4} data set over the SIR parameter space as a function of the error frequencies $\epsilon_N$ (a) and $\epsilon_T$ (b) for errors in node and time information, respectively. The curves are Levenberg--Marquardt fits to a stretched exponential form, $\omega_\infty[1-\exp(-a\epsilon^b)]$.}
\label{fig:diffmax_eml4}
\end{figure}

\begin{figure}
\includegraphics[width=0.65\columnwidth]{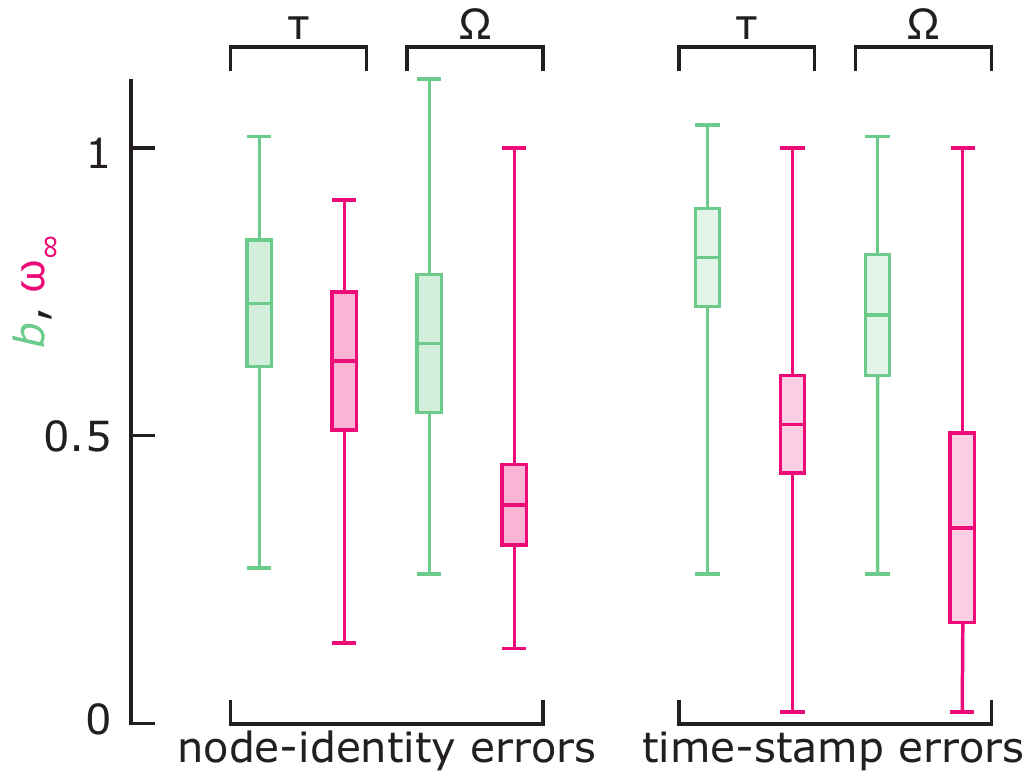}
\caption{ Box-and-whiskers plot of the fitting parameters $b$ and $\omega_\infty$. The box represents the region of one standard deviation. The other lines are the maximum, average and minimum respectively. $b$ is the stretching exponent; $\omega_\infty$ is the final value (in the limit of large errors).}
\label{fig:fitting_params}
\end{figure}

\section{Results}

\subsection{Example of the impact of node identity misinformation}

Turning to our numerical results, we first investigate the response to errors in the node identities for the \textit{E-mail 4} data set. (We choose this data set because it is of typical size and has all features we need to discuss below.) In Fig.~\ref{fig:eml4_n_xtme}(a), we show the deviation $\Delta\tau (\epsilon_N,\lambda,\delta)$ for an exponential progression of $\epsilon_N$-values---$\epsilon_N=10^{-3},10^{-2},10^{-1},1$---and the SIR parameters $\lambda$ and $\delta$.

As seen in Fig.~\ref{fig:eml4_n_xtme}(a), the response to the misinformation is nonlinear as functions of both $\epsilon$, $\lambda$ and $\delta$. For $\epsilon=10^{-3}$, the impact $\Delta\tau$ is less than $0.1$ throughout the $\lambda,\delta$-space. For $\epsilon=10^{-2}$, $\Delta\tau$ reaches values around $0.1$, while for larger $\epsilon_N$-values, $\Delta\tau$ is larger than $0.1$, or lower than $-0.1$ for a large part of parameter space. The shape of the region of large $\Delta\tau$ also changes with $\epsilon_N$. For $\epsilon_N\geq 0.1$ there are points with negative $\Delta\tau$. This region expands with $\epsilon_N$ originating from the large $\lambda,\delta$ limit. Fig.~\ref{fig:eml4_n_xtme}(b) shows how the parameter values $\lambda^\ast$ and $\delta^\ast$ maximizing $\Delta\tau$ changes with $\epsilon_N$. Indeed, neither of these quantities are constant---$\lambda^\ast$ and $\delta^\ast$ decreases ($\lambda^\ast$ much faster and more than  $\delta^\ast$). For most of the data sets, and both errors in time and node identities, $\lambda^\ast$ and $\delta^\ast$ is either constant or mostly decreasing---occasionally there can be local peaks but no increasing trend (plots not shown). To sketch an explanation for this behavior, first note that both randomizing time and network topology, with very few exceptions, facilitates spreading~\cite{holme_tempdis}. These phenomena have usually been attributed to heterogeneous distributions of inter-event times~\cite{Karsai2011} or network clustering~\cite{volz_cluster} respectively. Tuning up the error rates makes these phenomena increasingly strong. For small $\lambda$ and $\delta$ this means that the outbreak is more likely not to die out early and take hold in the population. This continues until a point where $\Delta\Omega$ and $\Delta\tau$ reaches their maxima after which $\Delta\tau$  decreases to negative values as the faster spreading makes the outbreak burn out fast in the population. For some data sets this peak happens close to $\lambda=1$, but for others it leads to a peak ~\cite{holme_extime_plos}.

\subsection{SIR parameter dependence of $\Delta$}

Now we will continue the analysis of $\Delta\Omega$ and $\Delta\tau$ on all the data sets and one value of the error frequency---$\epsilon_N=\epsilon_T=0.1$. In Figs.~\ref{fig:n_xtme} and \ref{fig:n_osze}, we show $\Delta\tau$ and $\Delta\Omega$ respectively for errors in node identities. We will not discuss every feature of these heat maps---just the most conspicuous---but also believe they give a feeling for the complexity, variability and universal features present in the data.

In general, the largest deviations happen for $\Delta\Omega$. $\Delta\Omega$ can reach up to $\Delta\Omega=0.75$ (see Fig.~\ref{fig:n_osze}). In this sense, prediction of outbreak sizes is more affected by topological misinformation than prediction of extinction times. The parameter dependence of $\Delta\tau$ falls into three distinct categories: a region of negative values at the region of large $\lambda$ and $\delta$ (\textit{Conference}, \textit{Hospital}, \textit{Office}, \textit{Primary School}, \textit{High School}, \textit{Kenya}, \textit{Diary}, \textit{WiFi}, \textit{Forum} and \textit{E-mail 1}, \textit{2} and \textit{3}); only a region of negative values at large $\delta$ (\textit{Romania} and \textit{UK}); and only a region of positive values (\textit{Gallery}, \textit{Facebook}, \textit{College} and \textit{E-mail 1}). Looking for possible structural explanations, \textit{Romania} and \textit{UK} are the data sets with highest number of contacts per node, while the latter group (including the \textit{Gallery} data) develop very slowly---this group e.g.\ tops the list of small $x_{nC}$ and $x_{lC}$ values. For large $\delta$, outbreaks are likely to burn out in the population, for datasets with many contacts per link this can happen also for quite low $\lambda$. Without misinformation there are typically new nodes coming into the data continuously which makes the outbreak last to the end of the sampling time, even though all nodes are affected. For the data sets with enough contacts per node, identity noise will make all nodes appear relatively early in the temporal network which will make the entire outbreak ending earlier, thus the pattern of \textit{Romania} and \textit{UK} in Fig.~\ref{fig:n_osze}. The phenomenon that low-$x_{nC}$-$x_{lC}$ networks have only positive $\Delta\tau$ can be understood since low $x_{nC}$ implies a big overturn of individuals. With a relatively high chance, the seed node would already have left the data by the time it gets infected which lowers $\Omega$. Node identity errors helps mixing the nodes in the time dimension too, which diminishes this  effect and increases the positive term of Eq.~\ref{eq:delta}.

The picture one gets from $\Delta\Omega$ (Fig.~\ref{fig:n_osze}) is slightly different from that of $\Delta\tau$ (Fig.~\ref{fig:n_xtme}). First $\Delta\Omega$ is non-negative, meaning that for no amount of noise, or location in the parameter space, $\Omega$ smaller than its original value. Node-identity noise thus always leads to an overestimation of the outbreak size. The second observation is that the peak of $\Delta\Omega$ is always at $\lambda=1$.  $\Omega$ is strictly increasing (both with and without misinformation), so it is not surprising that their difference also peaks at $\lambda=1$. Other than that, some patterns of Fig.~\ref{fig:n_osze} occurs in Fig.~\ref{fig:n_xtme} too---the very dense \textit{Romania} has large $\Delta\Omega$ deviations for low $\delta$ values while for large $\delta$, $\Omega$ is close to one for both the original and noisy networks (i.e.\ $\Delta\Omega$ is small).

In Figs.~\ref{fig:t_xtme} and \ref{fig:t_osze}, we plot the corresponding quantities to Figs.~\ref{fig:n_xtme} and \ref{fig:n_osze} but for noise in the time stamps rather than node identities. In Fig.~\ref{fig:t_xtme}, some data sets (\textit{Conference}, \textit{Hospital}, \textit{Office}, \textit{Primary School 1} and \textit{2}, \textit{High School 2}, \textit{Reality}, \textit{Romania}, \textit{WiFi}, \textit{UK} and \textit{E-mail 2}) have both regions of positive and negative $\Delta\tau$. \textit{WiFi} is different in that the region of negative $\Delta\tau$ lies below (lower $\delta$) the region of positive $\Delta\tau$. Compared to Fig.~\ref{fig:n_xtme} the data set \textit{Diary} is somewhat different in that it has $\Delta\tau\approx 0$ for errors in node identities but quite large $\Delta$ values for errors in the time stamps. In some of our network descriptors \textit{Diary} is extreme---it has the largest, or second largest, values of all the long-term activity quantities; it also has among the smallest values of the node and link inter-event-time burstiness. A node with small burstiness has relatively more uniformly spaced contacts. This mechanism is known to speed up spreading dynamics~\cite{Karsai2011}. Changing the node identities 
apparently affect these issues even more, otherwise expected, that in our paper from $[0.01,0.10]$.

As seen in Fig.~\ref{fig:t_osze}, $\Delta\Omega$ does not become negative when there are errors in the time of contacts either---no matter what the misinformation contains it always makes the predicted outbreaks larger. In a few cases however $\Delta\Omega$ is very low and close to zero (e.g.\ \textit{Gallery 1--5}, \textit{Prostitution}, \textit{Email 1}). The overall impression is similar to Fig.~\ref{fig:n_osze}, but for example \textit{Forum} is different with large $\Delta\Omega$ values for the node-identity errors but very small for time-stamp errors. \textit{Forum} is however not extreme in any of the structural measures that we use---it is one of the largest (in number of nodes, links and contacts) and one where the activity increases in the sense that $x_{nT}$ and $x_{lT}$ are among the smallest (then again \textit{Facebook} has even lower $x_{nT}$ and $x_{lT}$ but also smaller $\Delta\Omega$).

To sum up our results of the SIR parameter dependencies of the deviations $\Delta\tau$ and $\Delta\Omega$, there are some clear patterns, like the data sets sampled in the same way (\textit{Primary School}, \textit{High School} and \textit{Gallery}) are always showing the same patterns. On the other hand, the two major classes of data sets sampled---proximity networks and electronic communication do not show any consistent patterns. While authors have argued that social networks have a different structure than other networks~\cite{mejn:why}, we cannot say the same for these two classes.

\begin{table*}
\caption{\label{tab:linreg}Coefficient of determination $R^2$ and $p$-values of the multiple regression analysis of the relationship between the parameters describing $\omega(\epsilon)$ and network structural quantities. Stars represent three levels of significance (${}^{\ast\ast\ast}$ means $p<0.001$, ${}^{\ast\ast}$ means $0.001\leq p < 0.01$ and ${}^\ast$ means $0.01\leq p < 0.05$) and the absence of stars means $p \geq 0.05$.}
\begin{ruledtabular}
\begin{tabular}{l|dddddddd}
Structure &  \multicolumn{2}{d}{N\tau} & \multicolumn{2}{d}{N\Omega} & \multicolumn{2}{d}{T\tau} & \multicolumn{2}{d}{T\Omega} \\
 & b & \omega_\infty & b & \omega_\infty & b & \omega_\infty & b & \omega_\infty  \\ \hline
Long-term activity & 0.17 & 0.22 & 0.13 & 0.13 & 0.15 & 0.04 & 0.34^\ast & 0.24 \\
Node inter-event times & 0.25 & 0.67^{\ast\ast\ast} & 0.67^{\ast\ast\ast} & 0.03 & 0.22 & 0.01 & 0.38 & 0.08 \\
Node duration & 0.16 & 0.04 & 0.07 & 0.60^{\ast\ast\ast} & 0.26 & 0.01 &  0.33^\ast & 0.25 \\
Node activity &0.20 & 0.75^{\ast\ast\ast}  & 0.64^{\ast\ast\ast} & 0.50^{\ast\ast} & 0.30 & 0.02 & 0.44^{\ast\ast} & 0.18 \\
Link inter-event times & 0.26 & 0.41^{\ast\ast} & 0.64^{\ast\ast\ast} & 0.02 & 0.23 & 0.01 & 0.39^{\ast\ast} & 0.05 \\
Link duration & 0.13 & 0.06 & 0.10 & 0.13 & 0.24 & 0.02 & 0.22 & 0.25 \\
Link activity & 0.22 & 0.76^{\ast\ast\ast} & 0.66^{\ast\ast\ast} & 0.36 & 0.21 & 0.14 & 0.35^\ast & 0.14 \\
Full nwk.\ deg.\ dist. & 0.22 & 0.04 & 0.07 & 0.82^{\ast\ast\ast} & 0.33^\ast & 0.26 & 0.10 & 0.15 \\
Full nwk.\ structure & 0.27 & 0.16 & 0.24 & 0.23 & 0.27 & 0.12 & 0.38^\ast & 0.31^\ast \\
Red.\ nwk.\ deg.\ dist. & 0.05 & 0.08 & 0.07 & 0.81^{\ast\ast\ast} & 0.13 & 0.18 & 0.24 & 0.24 \\
Red.\ nwk.\ structure & 0.08 & 0.30^\ast & 0.23 & 0.16 & 0.26 & 0.18 & 0.33^\ast & 0.19 \\
System sizes & 0.17 & 0.12 & 0.53^{\ast\ast\ast} & 0.45^{\ast\ast\ast} & 0.32 & 0.01 & 0.42^{\ast\ast} & 0.19 \\
\end{tabular}
\end{ruledtabular}
\end{table*}

\subsection{Impact of error rate on prediction deviations}

To better understand the response of the level of misinformation on the prediction accuracy, we study $\omega(\epsilon)$---the maximum absolute value of $\Delta\tau$ or $\Delta\Omega$-values (see Eq.~(\ref{eq:epsilon})). We searched for a functional form that can summarize the $\epsilon$-dependence of $\omega$---not necessarily the best statistical model in a model selection sense, but a function that fit all misinformation scenarios. We found a stretched exponential convergence---
\begin{equation}\label{eq:stretched}
\widetilde{\omega}(\epsilon)=\omega_\infty\left[1-\exp(-a\epsilon^b)\right] ,
\end{equation}
where $a$ and $b$ are fitting parameters---to meet this condition very well. None of the scenarios for any of the networks has a reduced $\chi^2$ value larger than $1$ (it ranges from $5.3\times 10^{-8}$ for \textit{Kenya} $\omega_\Omega$ of time errors to $0.24$ for the $\omega_\tau$ of node identity errors of \textit{E-mail 1}). The parameter $b$ (typically in the interval $0<b<1$) is called the \textit{stretching exponent} and its deviation from unity indicates how much the tail is stretched compared to an exponential decay~\cite{sornette}. An example of this behavior---neither the best, nor the worst fit---can be seen in Fig.~\ref{fig:diffmax_eml4} (for the \textit{E-mail 4} dataset---the same example as in Fig.~\ref{fig:eml4_n_xtme}). 

As far as we understand, there is no straightforward explanation for this functional form. Rather, we believe that in general the $\omega(\epsilon)$-curves can deviate from stretched exponentials. Indeed, the points that are off the fitting curves (e.g.\ the point $\omega_\tau(\epsilon_N=0.004)$) are probably not a result of bad convergence, but structures in the data sets. The three fitting parameters of Eq.~\ref{eq:stretched} are nevertheless concise ways of summarizing the shapes of the $\omega(\epsilon)$ dependence and a way of relating temporal network structure and the error response of epidemic predictions.

As alluded to, the perhaps most interesting parameter of the stretched exponential fits is the stretching exponent $b$. If $b=1$, the convergence is exponential. If $b<1$, the decay is stretched (or slower than exponential). In Fig.~\ref{fig:fitting_params}, we present the values for our eight analyses (we refer to them as ``cases'' below)---$b$ or $\omega_\infty$, node-identity or time-stamp misinformation, or prediction of $\tau$ or $\Omega$. It is almost the case for all misinformation scenarios and data sets that $0<b<1$ (the maximal observed $b$-value is $1.2$). Otherwise, we note that there is a fairly large spread of $b$, and that it is quite similar for all combinations of outbreak descriptor ($\tau$ or $\Omega$) and misinformation type. $\omega_\infty$ is bounded to $[0,1]$ and does indeed take values in the entire range. The average $\omega_\infty$ node-identity misinformation and $\Delta\tau$ is larger than the others which is also hinted from Figs.~\ref{fig:n_xtme}--\ref{fig:t_osze}.

As a final analysis, we seek to understand the values of the fitting parameters $b$ and $\omega_\infty$ for the individual data sets (not only the summary statistics of Fig.~\ref{fig:fitting_params}). We will once again look for explanations in our network descriptors. To do this in a systematic way, we perform an independent multiple regression analysis for each structural category listed in Section~\ref{net_desc} (quantities related to the long-term activity, the node inter-event time, etc.). We fit a linear model of the variables of each category to the $b$ and $\omega_\infty$ values. To keep the analysis simple we do not include interaction variables (that would be interesting for the future). Table~\ref{tab:linreg} shows the $R^2$ values (coefficients of determination) of the regression analysis for each combination of parameters $b$ (or $w_{\infty}$) and scenarios/measures of misinformation, i.e.\ $N\tau$ (node-identities and extinction time), $N\Omega$ (node-identities and outbreak size), $T\tau$ (time-stamps and extinction time) and $N\Omega$ (time-stamps and outbreak size). It also marks the p-values (giving significance levels) of the hypothesis that there is no correlation between the temporal network structure and $b$ (or $w_{\infty}$).

This analysis shows that none of our classes of temporal-network quantities we are investigating is a very strong predictor of the deviations---$R^2$ is never very close to $1$. On the other hand, there is nothing else than the temporal network structure that affects the outbreaks, so we cannot conclude the study of network-structural causes of the sensitivity to misinformation. On the other hand, there are statistically significant correlations, and they depend on the quantity to be predicted and the misinformation scenarios. Three cases can be well-described by the data---$\omega_\infty$ for the extinction time measurements with node-identity misinformation ($N\tau$) and both $b$ and $\omega_\infty$ for the outbreak sizes of the same misinformation scenario ($N\Omega$). In all of these cases, the node activity distributions is a strong predictor. Node and link inter-event times are also important factors determining $b$ for the $N\Omega$ case and $\omega_\infty$ for $N\tau$. For $\omega_\infty$ and $N\Omega$, the degree distributions and system sizes seem more important. It is hard to have some further intuition in why different responses to misinformation is differently affected by temporal-network structure. There is, seemingly, no intermediate connection between temporal descriptors (like inter-event times) and temporal misinformation, or our temporal outbreak characteristic (the extinction time).

\section{Discussion}

We have studied the effects of misinformation in contact data on the predictability of epidemics by the SIR model. We find a complex situation where both misinformation in time stamps and node identities affect the prediction of both outbreak sizes and times to extinction in fairly similar orders of magnitude. The maximal deviation between predictions from the original and erroneous data follows a stretched exponential convergence as a function of the error frequency.

The variation of our results across data sets are very large, but the parameter dependence fall into classes that can be understood to some extent from the particular way they were sampled. For the proximity networks, the definition of contacts naturally affects the density of contacts. We did not try to normalize the datasets as this can be done by adjusting the SIR parameters. For the future, it would be interesting to know if the results could be rescaled to collapse the deviation heat-maps (e.g.\ Fig.~\ref{fig:n_xtme}).

We, furthermore, examine what kind of simple temporal-network quantities that controls the response to misinformation. We find that this question depends on the type of misinformation and the type of prediction. For example, the stretching exponent of how the maximal deviation of outbreak sizes depend on the frequency of node-identity errors is influenced by the node and link inter-event time distributions. (The stretching exponent is positively correlated with the mean and standard deviation of the inter-event times, but negatively correlated with the coefficient of variation and skewness.) In general, inter-event times and degree distributions seem more influential than the long-term time activity of the data sets.

More generally speaking than the response to misinformation, what temporal structures that are the most important determinants for disease spreading in temporal networks is currently a matter of debate. Ref.~\cite{holme_tempdis} compares three levels of representing contact networks (temporal, static and fully-connected networks) and argues that the long-term time activity best explains how well SIR disease prediction on these representations match. It is hard to compare to our results since at no parameter values do our randomization not match the representations that Ref.~\cite{holme_tempdis} is comparing. Is is possible that long-term time activity is more important for what representation to choose, while other structures dominate the response to misinformation. Furthermore, Ref.~\cite{holme_liljeros} shows that when coarse graining a temporal network, keeping the underlying network fixed, it is more important to keep the time evolution than the inter-event times. As both node-identity and time-stamp misinformation affects both the inter-event times and the long-term activity, this is still consistent with our work. Still, it is remarkable that it seems like not some types of temporal network structure that we test seem to have more explanatory power than others---they all seem to matter but different ones being more important for different issues. The most important conclusion is that it can be deceiving to rely on models tuning simple temporal-network structures, say the degree distribution, to evaluate intervention methods (like network vaccination, etc.)---it is simply hard to say what structure that matters in what situation.

The quest for understanding how temporal-network structure influences disease spreading continues. One interesting extension of the approach in this paper would be to---rather than looking at global quantities such as $\omega$, $\Delta\tau$, etc., to look at different nodes or regions of the same data and the behavior of these. If one changes the contact patterns of individual nodes or clusters, how does that change the disease propagation?

\bibliographystyle{abbrv}
\bibliography{misinfo}

\end{document}